\documentclass[10pt]{article}

\usepackage{cite}
\usepackage{graphicx}        
\usepackage{amssymb}
\usepackage{amsmath}
\usepackage{pdflscape}
\usepackage{tikz}
\usepackage[scriptsize]{subfigure}
\usepackage{amsmath}
\usepackage{soul}

\usepackage{float}

\usepackage[T1]{fontenc}
\usepackage{kpfonts,baskervald}

\usepackage[utf8]{inputenc}
\usepackage{authblk}

\usepackage{placeins}

\usepackage{caption} 
\usepackage{dblfloatfix} 

\tikzset{cross/.style={cross out, draw=black, minimum size=2*(#1-\pgflinewidth), inner sep=0pt, outer sep=0pt},
cross/.default={2.5pt}}
\usepackage[margin=1in]{geometry}
\usepackage{parskip}

\DeclareGraphicsExtensions{.pdf,.png,.jpg,.eps} 
\usepackage{epstopdf}
\usepackage[acronym,shortcuts]{glossaries}

\makeglossaries

\newacronym{GFVSC}{\textcolor{black}{GF-VSC}}{Grid-forming voltage source converter}

\begin{document}
%
\title{Fast voltage boosters to improve transient stability of power systems with 100\% of grid-forming VSC-based generation}
\date{ }

\author[1]{R\'egulo E. \'Avila-Mart\'inez} \author[1]{Javier Renedo} \author[1]{Luis Rouco} \author[1]{Aurelio Garc\'ia-Cerrada} \author[1]{Lukas Sigrist} \author[2]{Taoufik Qoria} \author[2]{Xavier Guillaud}

\affil[1]{Instituto de Investigaci\'on Tecnol\'ogica (IIT), ETSI ICAI, Universidad Pontificia Comillas,  Madrid, Spain}
\affil[2]{Laboratoire d'Electrotechniquede Puissance (L2EP), Centrale Lille,
Arts et M\'etiers Paris Tech, University of Lille, Lille, France}

\maketitle

\section*{Abstract}\label{sec:abstract}
\ac{GFVSC} has been identified as the key technology for the operation of future converter-dominated power systems. Among many other issues, transient stability of this type of power systems remains an open topic of research because it is still a key limiting factor for stressed power systems. Previous studies have proposed control strategies for GF-VSC to improve transient stability of this type of systems by suitable current-limitation algorithms and/or control of active-power injections. As an alternative, this paper proposes two fast voltage boosters to improve transient stability of power systems with 100~\% of GF-VSC-based generation with virtual synchronous machine~(VSM). One control strategy uses local measurements, whereas the other one uses global measurements of the frequency of the centre of inertia (COI). Both strategies improve transient stability of this type of systems significantly. The advantage of using fast voltage boosters for this purpose is that the set points linked to frequency/active-power injection~(i.e set points linked to the primary energy source of the VSCs) will not be modified. Furthermore, strategies such as current-limitation, active-power control and fast voltage controllers for transient stability improvement are compatible and complementary.

\noindent \textbf{Index terms:} Voltage source converter, VSC, grid forming, transient stability


\noindent This is an unabridged draft of the following  paper (submitted to IEEE Transactions on Smart Grid on 4-6-2021):
\begin{itemize}
	\item R. E. \'Avila-Mart\'inez, J. Renedo, L. Rouco, A. Garc\'ia-Cerrada, L. Sigrist, T. Qoria, X. Guillaud, \emph{``Fast voltage boosters to improve transient stability of power systems with 100\% of grid-forming VSC-based generation"}, Manuscript ID: TSG-00871-2021, pp. 1-8, 2021. 
	\item Internal reference of this paper: IIT-21-078WP.
\end{itemize}

\newpage

\vspace{-0.3cm}
\section{Introduction}\label{sec:intro}
\noindent Future power systems are planned for massive integration of renewable energy sources. Non-conventional renewable generators, such as wind turbines and solar photovoltaic (PV) generators are interfaced through voltage source converters~(VSC). Grid-forming VSC (\ac{GFVSC})  has been identified as the key technology for the operation of future converter-dominated power systems, from small microgrids~\cite{Rocabert2012,Guerrero2013,Olivares2014} to large transmission systems~\cite{Milano_lowH2018,Paolone2020}. The main characteristic of a \ac{GFVSC} (also known as voltage-controlled VSC) is that it controls its output voltage magnitude and frequency and therefore, it is capable of creating a grid. This is not possible with most extended grid-following VSCs (also known as current-controlled VSCs or grid-feeding VSCs). When operating a power system with several \acp{GFVSC}, they need a self-synchronisation mechanism, in order to ensure that all of them reach the same frequency. Fortunately, unlike in grid-following VSCs, synchronisation can be achieved without using a phase-looked loop (PLL) by means of supplementary control strategies. Although there are variants of these supplementary control strategies, the philosophy behind most of them is that they control \acp{GFVSC} in order to mimic the behaviour of conventional synchronous generators. The most common option in alternating-current (AC) microgrids is the so-called power-frequency (P-f) droop~\cite{Rocabert2012,Guerrero2013,Olivares2014}, while the most common option in large power systems is the emulation of synchronous machines, in the so-called syncrhonverters~\cite{Zhong2011} or virtual synchronous machines (VSM)~\cite{DArco2015,jroldan2019}. Furthermore, the work in~\cite{DArco2014} proved that P-f droop supplementary controllers and VSM supplementary controllers are equivalent. A different approach for self-synchronisation of \acp{GFVSC} is the concept of reactive-power synchronisation, proposed recently in~\cite{Amenedo2021}.

Transient stability (or angle stability with large disturbances) is defined as the ability of the generators of the system to remain in synchronism when large disturbances occur~\cite{Nikos2020}. Initially, this definition assumed that the system contains synchronous generators, since the phenomenon is related to the rotor angle of synchronous machines. However, recent publications have shown that transient stability is also a concern in power systems with 100\% of \ac{GFVSC}-based generation. In other words, if \acp{GFVSC} emulate synchronous machines, they can also lose synchronism in case of severe-enough faults. 

The work in~\cite{Andrade2011,Andrade2011b} observed that the loss-of-synchronism phenomenon can occur in AC microgrids with \ac{GFVSC} generators equipped with P-f droop control; transient stability was assessed by means of Lyapunov's theory. Further studies are presented in~\cite{Xin2016,Eskandari2020}, where the impact of current saturation~\cite{Xin2016} and fault-ride-through~(FRT) capability~\cite{Eskandari2020} of \acp{GFVSC} on transient stability is analysed. References~\cite{Shuai2018, Cheng2020} analysed transient stability in power systems with \ac{GFVSC} controlled as VSMs. The study in~\cite{Cheng2020} included experimental results. A similar approach is followed in~\cite{Pan2020}, where a thorough analysis of transient stability of \acp{GFVSC} is provided, and it also includes experimental results. The work in~\cite{Qoria_VSC_current_limit2020} analyses the impact of current-limiting strategies on transient stability of a \ac{GFVSC} connected to an infinite grid. The study analyses two current-limiting strategies: the conventional current limiter for the current modulus in vector control and virtual-impedance-based current limiter. Eventually, the work proposes an hybrid current limiter which is a combination of both, improving the performance of the current limitation process as well as transient stability. The work in~\cite{Qoria_VSC_CCT2020} goes a step further and analyses transient stability of a \ac{GFVSC} connected to an infinite grid, using virtual-impedance current limiters of the \ac{GFVSC}. The work proposes an adaptative droop in the \ac{GFVSC} by changing the proportional gain of the P-f droop controller during the fault. Results show that the strategy increases the crititical clearing time (CCT) of a fault, significantly. The work in~\cite{Qoria2020} proposed an adaptive emmulated inertia in the \ac{GFVSC} in order to improve transient stability. The three studies~\cite{Qoria_VSC_current_limit2020,Qoria_VSC_CCT2020,Qoria2020} include experimental results. Reference~\cite{Shen2020} analysed the case of a \ac{GFVSC} (with VSM) and a grid-following VSC connected to an infinite grid. The study proposed a control strategy for the grid-following VSC for transient stability disabling the PLL of the grid-following VSC during the fault and using, instead, remote measurements to obtain the virtual rotor angle angle of the \ac{GFVSC} and use it for vector control of the grid-following VSC. The work in~\cite{Choopani2020} analysed transient stability of a power system with 100\% of \ac{GFVSC}-based generators working like VSMs. The work proposed an active-power~(P) control strategy based on a PI controller using, as input signal, the frequency deviation with respect to the frequency of the centre of inertia (COI), in order to improve transient stability and producing promising results. The work in~\cite{Xiong2021} proposes a supplementary P-control strategy proportional to the frequency deviation with respect to the nominal frequency (using local measurements) to improve transient stability of a \ac{GFVSC} connected to an infinite grid. The study provides guidelines for the design of the proposed controller and it includes experimental results.

The research studies described above have shown that transient stability in power systems with \acp{GFVSC} can be improved (a) by implementing suitable current limiters~\cite{Qoria_VSC_current_limit2020,Qoria_VSC_CCT2020} or (b) by implementing suitable active-power-related control strategies~\cite{Qoria_VSC_current_limit2020,Qoria_VSC_CCT2020,Qoria2020,Choopani2020}. However, the use of control strategies based on reactive-power~(Q) injections or voltage control to improve transient stability in power systems with 100~\% of \ac{GFVSC}-based generation has not been investigated in previous work. Furthermore, the use of reactive-power/voltage control strategies to improve transient stability have been successfully applied in traditional synchronous generators (i.e. excitation boosters)~\cite{Lee1986,Kundur1994,LuisDM2016b,LuisDM2017,LuisDM2019,LuisDM2020}, in shunt FACTS devices~\cite{Haque2004a} and in high voltage direct current systems based on voltage source converters~(VSC-HVDC)~\cite{Fuchs2014,iitcontrolQ2017}. The good results obtained there, has motivated this research.

An important advantage of controlling voltage at the output of the filter of the \ac{GFVSC}  is that the converter will change its reactive-power injection, while the set points linked to frequency/active-power injection will not be modified. In other words, this type of controllers do not involve changing the set point of the primary energy source of the \ac{GFVSC}. Along this line, this work proposes two control strategies based on fast voltage boosters to improve transient stability in power systems with 100\% of \ac{GFVSC}-based generation. One control strategy uses local mesurements, whereas the other one uses global measurements of the frequency of the COI. Both strategies improve transient stability of this type of systems significantly.

Specifically, the contributions of this work are as follows:
\begin{itemize}
	\item Proposal of two fast voltage boosters to improve transient stability of power systems with 100\% of \ac{GFVSC}-based generation, one based on local measurements and the other one based on global measurements.
	\item Significant improvements on the critical clearing times of different faults, thanks to the two control strategies.
	\item Analysis of the impact of communication latency on the performance of the proposed global control strategy. Results show that the proposed control strategy is robust for realistic communication latency. 
	\item Discussion about the use of local and global measurements in this type of controllers.
\end{itemize}



\section{Grid-forming VSCs}\label{sec:VSC_V}
\subsection{Modelling and control}\label{sec:VSC_V_model}

\begin{figure}[!htbp]
\begin{center}
\includegraphics[width=0.75\columnwidth]{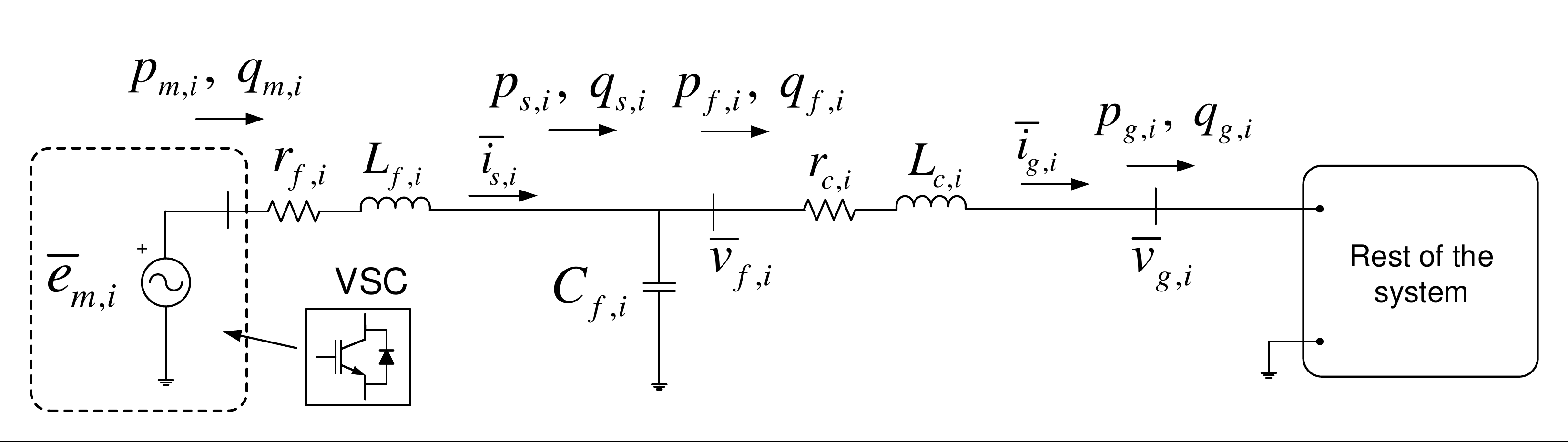}
\caption{Model of a grid-forming VSC. }
\label{fig:VSC_V_model}
\end{center}
\end{figure}

\noindent This section describes the model of a \ac{GFVSC}, following the guidelines of~\cite{Qoria2018,Paolone2020,Rokrok2020,Qoria2020,Pereira2020}. Fig.~\ref{fig:VSC_V_model} shows the equivalent model of a \ac{GFVSC}-$i$. The converter is represented as a voltage source~($\bar{e}_{m,i}$) and it is connected to the rest of the system through an LC filter, consisting of a phase reactor ($\bar{z}_{f,i}=r_{f,i} + j \omega_{i} L_{f,i}$) and a capacitor ($\bar{z}_{Cf,i}=-j/(\omega_{i} C_{f,i})$), and a transformer ($\bar{z}_{c,i}=r_{c,i} + j \omega_{i} L_{c,i}$). 

\begin{figure}[!htbp]
\begin{center}
\includegraphics[width=1.00\columnwidth]{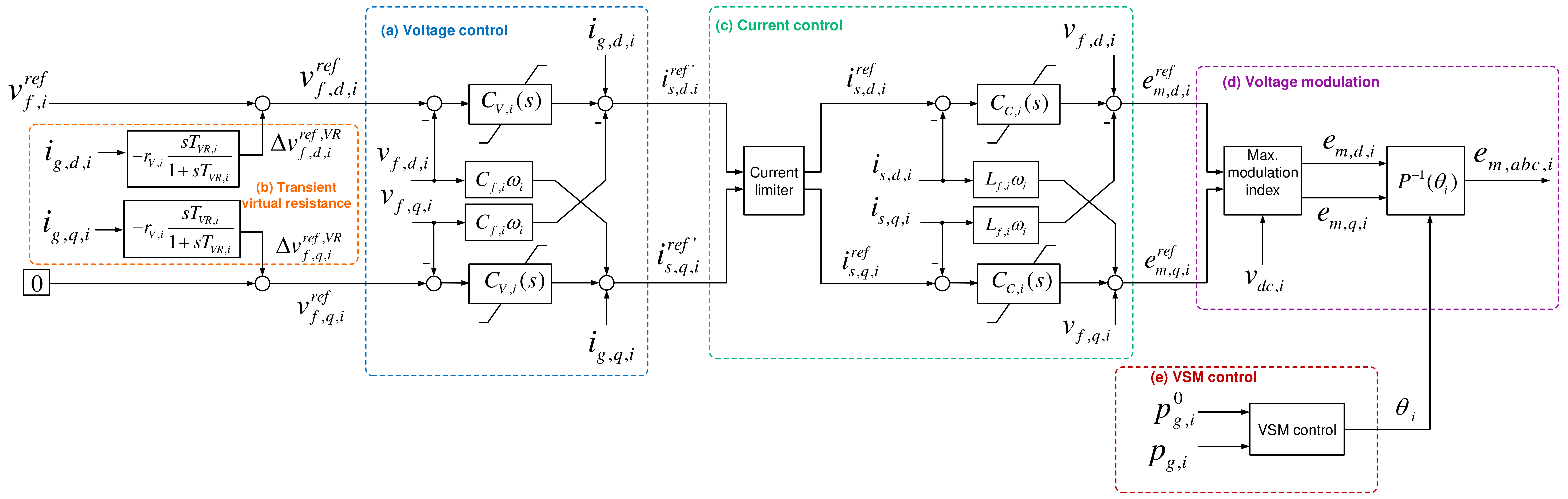}
\caption{General scheme of the control system of a grid-forming VSC. }
\label{fig:VSC_V_control_loops_general}
\end{center}
\end{figure}

The \ac{GFVSC}-$i$ will control:
\begin{itemize}
	\item The magnitute of the voltage at bus $f$: $v_{f,i}$.
	\item The frequency at bus $f$: $\omega_{f,i}$. This means that the VSC controls the  angle of the voltage at bus $f$: $\delta_{f,i}$. 
\end{itemize}

Fig.~\ref{fig:VSC_V_control_loops_general} shows the general control structure of a \ac{GFVSC}-$i$ based on vector control~\cite{Rokrok2020,Qoria2020}. It consists of: (a) a voltage controller, (b) a virtual transient resistance, (c) a current controller, (d) voltage modulation and (e) a grid-forming mechanism for self-synchronisation (e.g. VSM or any other variant). The details of the control system can be found in~\cite{Qoria2018,Paolone2020,Rokrok2020,Qoria2020,Pereira2020}. Hence, only a summary is provided here.

\begin{figure}[!htbp]
\begin{center}
\includegraphics[width=0.60\columnwidth]{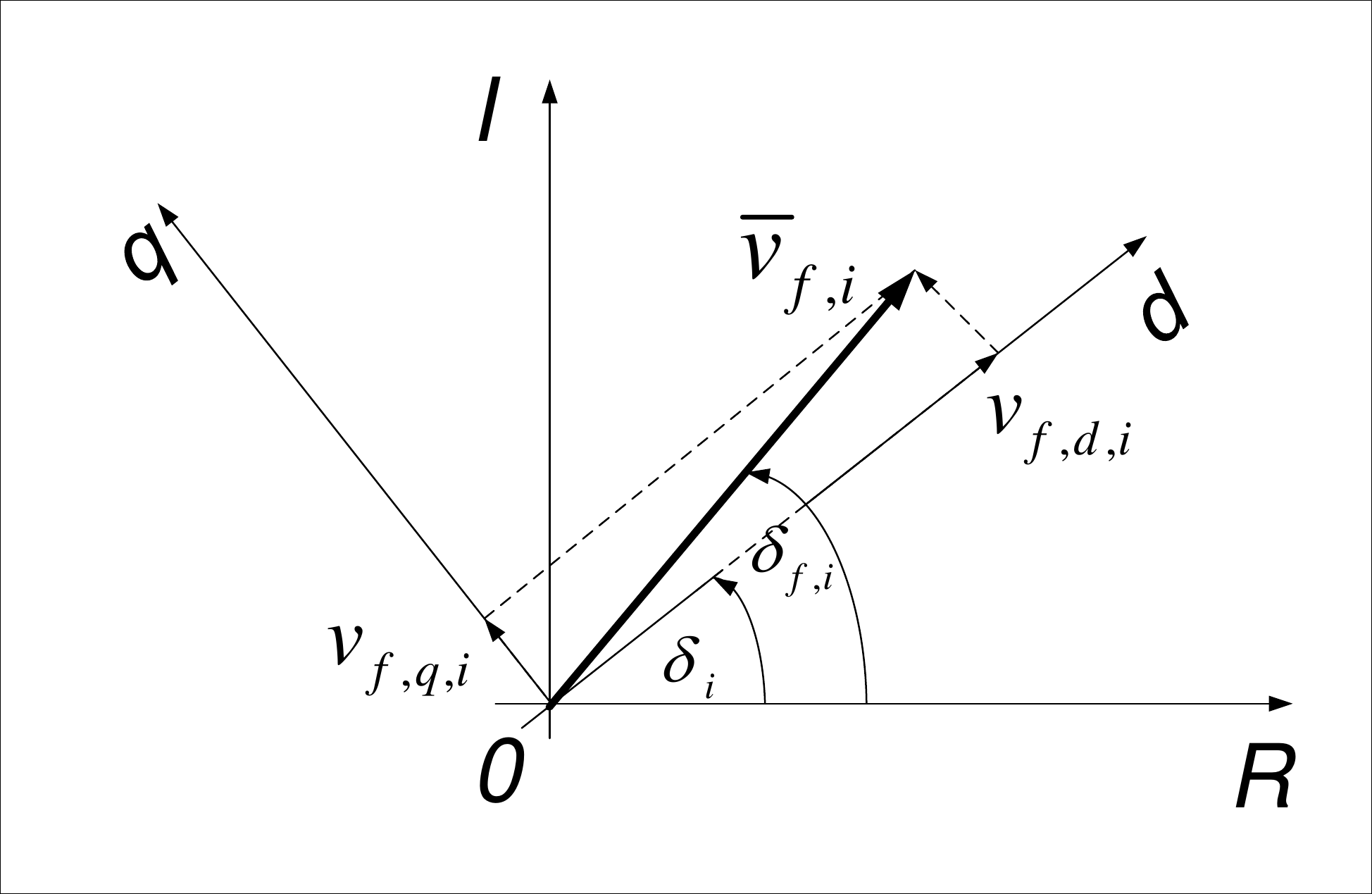}
\caption{Grid-forming VSC: $d-q$ axes. }
\label{fig:VSC_GM_dq_axes_v1}
\end{center}
\end{figure}

First of all, the VSC imposes an angle $\delta_i$ (rad) for the mobile $d-q$ reference frame which rotates at a speed $\omega_{i}$ (pu), as illustrated in Fig.~\ref{fig:VSC_GM_dq_axes_v1}. The angle $\delta_i$ is referred to an arbitrary mobile $R-I$ reference frame rotating at the synchronous frequency. The position of the $d$ axis with respect to a static reference frame (angular speed equal zero), $\theta_{i}$,  is the one used in Park’s Transform to refer variables to the $d-q$ reference frame. The way in which this angle is obtained from the self-synchronisation method (VSM control in Fig.~\ref{fig:VSC_V_control_loops_general}) will be described in Section~\ref{sec:VSC_V_VSM}.

The voltage control loop (Fig.~\ref{fig:VSC_V_control_loops_general}) controls $v_{f,d,i}$ and $v_{f,q,i}$ with proportional-intergral controllers (PI) where set-point values are given by:
\begin{eqnarray}\label{eq_VSC_GM_out_Vf}
	v_{f,d,i}^{ref}& = & v_{f,i}^{ref} + \Delta v_{f,d,i}^{ref,VR} \\ \nonumber
	v_{f,q,i}^{ref}& = & 0 + \Delta v_{f,q,i}^{ref,VR} 
\end{eqnarray}
where $v_{f,i}^{ref}$ is the set point of the magnitude of the voltage and {$v_{f,q,i}^{ref}$} is controlled to zero. $\Delta v_{f,d,i}^{ref,VR}$ and $\Delta v_{f,q,i}^{ref,VR}$ represent the transient virtual resistance: they only have impact during transients and are used for damping~\cite{Rokrok2020}. Eventually, the voltage controller aligns the voltage $\bar{v}_{f,i}$ with the $d-q$ axes of the VSC (see Fig.~\ref{fig:VSC_GM_dq_axes_v1}). 

The outputs of the voltage controller are the set points of $d-q$ current components of the current controllers, which have a current limiter and PI controllers. The current limiter is implemented with the conventional current saturation algorithm used in vector control~\cite{Qoria_VSC_current_limit2020} (i.e. the modulus of the current vector is limited). The outputs of the current controllers are the modulated voltages $e_{m,d,i}$ and $e_{m,q,i}$, related to the DC voltage as:
\begin{equation}\label{eq:VSC_GF_em}
\bar{e}_{m,i}= \bar{m}_{i} v_{dc,i}, \mbox{\space \space with  \space} e_{m,i} \leq m_{i}^{max} v_{dc,i} \mbox{ (pu).}
\end{equation}
where $\bar{m}_{i}$ is the modulation index. The maximum modulation index (magnitude), in pu, can be calculated as:
\begin{equation}\label{eq:m_max_pu}
	m_{i}^{max} = \sqrt{\frac{3}{2}} \cdot \frac{V_{dc,B}}{2 V_{ac,B}} \mbox{\space (pu)}
\end{equation}
where $V_{dc,B}$ is the DC -voltage base value (pole to pole) and $V_{ac,B}$ is the AC-voltage base value (phase to phase).

\subsection{Virtual synchronous machine control (VSM)}\label{sec:VSC_V_VSM}
\begin{figure}[!htbp]
\begin{center}
\includegraphics[width=0.85\columnwidth]{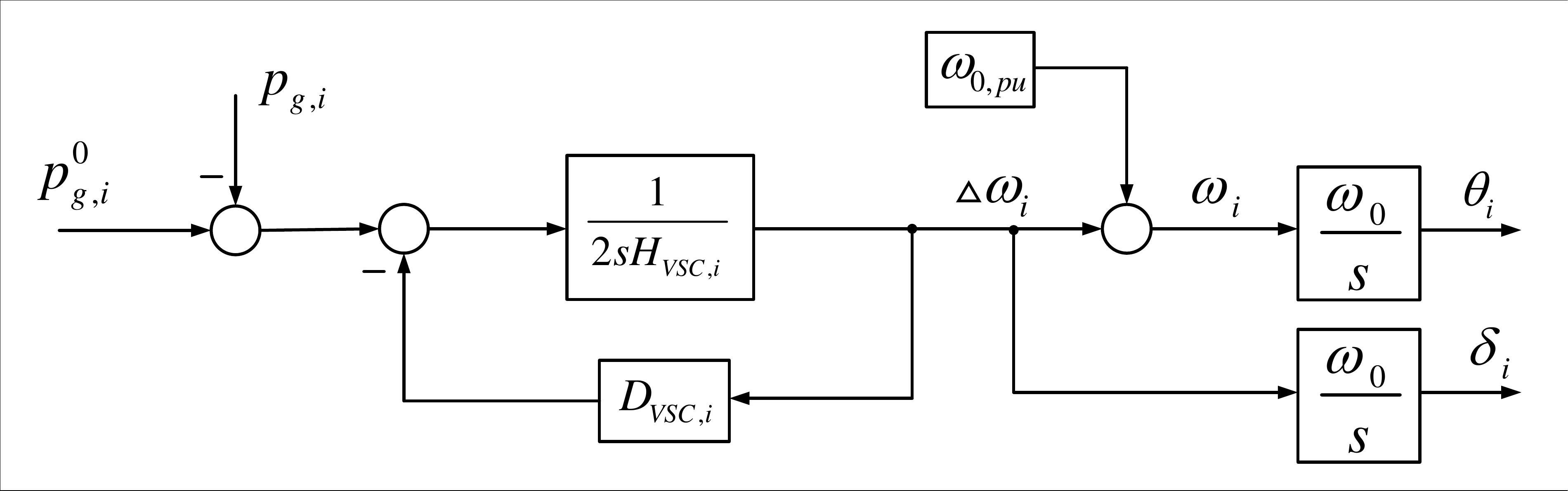}
\caption{Supplementary controller of a virtual synchronous machine (VSM) implemented in a \ac{GFVSC}. }
\label{fig:VSC_V_VSM}
\end{center}
\end{figure}

\noindent A virtual synchronous machine (VSM) is implemented as a supplementary controller that manipulates the set points of the outer controllers of the \ac{GFVSC}. Guidelines for the implementation of VSM controllers and different implementations can be found in~\cite{DArco2015,Paolone2020,Choopani2020,Rokrok2020,Qoria2020}. This work considers a VSM supplementary controller amulating a classical model of a synchronous machine and equipped with a primary frequency controller, as shown in Fig.~\ref{fig:VSC_V_VSM}. 

The VSM emulates the swing equation of a synchronous machine:
\begin{eqnarray}\label{eq_VSC_V_VSM_v1}
	p_{g,i}^{0} - p_{g,i} -D_{VSC,i} \Delta \omega_{i} & = & 2 H_{VSC,i}\frac{d \Delta \omega_{i}}{dt}  
\end{eqnarray}
where:
\begin{itemize}
	\item $H_{VSC,i}$~(s) is the emulated inertia constant.
	\item $D_{VSC,i}=1/R_{PFR,i}$~(pu) is the proportional gain of the primary frequency controller~(or primary frequency response, PFR). It is the inverse of the primary frequency droop constant ($R_{PFR,i}$). 
	\item $\Delta \omega_{i}=\omega_{i}-\omega_{0,pu}$~(pu) is the frequency increment of the \ac{GFVSC} ouput frequency, $\omega_{i}$, with respect to the nominal frequency, $\omega_{0,pu}=1$~pu. 
	\item $p_{g,i}^{0}$~(pu) is a constant active-power set point of the \ac{GFVSC} at the PCC. This term represents the emulated mechanical power of the VSM.
	\item $p_{g,i}$~(pu) is the  active-power delivered by the \ac{GFVSC} measured at the PCC.
	\item $\omega_0$ is the nominal frequency in rad/s.
\end{itemize}
Summarising, eventually \ac{GFVSC}-$i$ of~Fig.~\ref{fig:VSC_V_model} will impose the frequency given by the swing equation~(Fig.~\ref{fig:VSC_V_VSM} and~(\ref{eq_VSC_V_VSM_v1})): $\omega_{i}$~(pu). The emulated rotor angle, $\delta_{i}$~(rad), is the angle of the $d-q$ reference frame (Fig.~\ref{fig:VSC_GM_dq_axes_v1}) and $\theta_{i}$~(rad) is the angle used for Park's transformation in Fig.~\ref{fig:VSC_V_control_loops_general}. It should be remarked that~(\ref{eq_VSC_V_VSM_v1}) is a control algorithm imposed by the VSC.

\FloatBarrier

\section{Proposed fast voltage boosters to improve transient stability}\label{sec:VSC_V_TS}
\noindent In this section, fast voltage boosters for \acp{GFVSC} are proposed to improve transient stability. The proposed controllers are based on reactive-power/voltage control in the VSCs and they were inspired by the excitation boosters  for synchronous machines proposed in~\cite{LuisDM2017,LuisDM2019,LuisDM2020} to improve transient stability in multi-machine systems. 

The active-power injection of the VSC at the connection point (see Fig.~\ref{fig:VSC_V_model}) can be approximated as:
\begin{equation}\label{eq:VSC_pg_electrical}
p_{g,i} \simeq \frac{v_{f,i} v_{g,i}}{x_{c,i}} \sin (\delta_{f,i}-\delta_{g,i})
\end{equation}
Eq.~(\ref{eq:VSC_pg_electrical}) is useful for qualitative analysis. Loss-of-synchronism in power systems with 100~\% of \ac{GFVSC}-based generation follows the same pattern as in multi-machine systems with synchronous generators. When a fault occurs, the voltage at the PCC, $v_{g,i}$, is reduced dramatically, reducing the active power injection $p_{g,i}$ (i.e. the virtual electromagnetic torque of the VSC). This produces the acceleration of the VSCs (see the (\ref{eq_VSC_V_VSM_v1})). Depending on the location of the fault, some VSCs will accelerate more than others and severe-enough faults could produce loss of synchronism of \acp{GFVSC}.

Equation~(\ref{eq:VSC_pg_electrical}), shows that the active-power injection can be modified by changing voltage $v_{f,i}$ and this is feasible, since voltage controllers of \acp{GFVSC} have fast responses. Therefore, the voltage set point of each \ac{GFVSC}-$i$ can be modified with an additional term ($\Delta v_{f,i}^{ref,TS}$) seeking the improvement of transient stability:
\begin{equation}\label{eq:VSC_vf_ref_TS}
v_{f,i}^{ref} = v_{f,i}^{0}  + \Delta v_{f,i}^{ref,TS}
\end{equation}
where $v_{f,i}^{0}$ is the initial voltage set point. Notice that voltage set point $v_{f,i}^{ref}$ is actually the input of the voltage controller (\ref{eq_VSC_GM_out_Vf}) (see Fig.~\ref{fig:VSC_V_control_loops_general}).

A control strategy will be effective to improve transient stability if it is able to act during the fault and/or inmediately after the fault clearing. In fact, VSCs might not be able to produce any effect during the fault, because the converter will limit its current injection, if the fault is close enough.

Transient stability of a single \ac{GFVSC} connected to the grid can be improved by slowing down the VSC, following a pattern similar to the one used for a single synchronous machine connected to an infinite grid~\cite{Kundur1994,LuisDM2016b}. Transient stability of a power system with 100~\% of \ac{GFVSC}-based generation is a more complex phenomenon. A fault will produce that some VSCs accelerate faster than others during the transient and control actions should try to pull their frequency together. In other words, some VSCs will have to be slowed down while others will have to be accelerated. This is, again, analogous to the case of a multi-machine system with conventional synchronous generators, where the use of the speed of the centre of inertia (COI) has proved to be useful~\cite{LuisDM2017,LuisDM2019,LuisDM2020}.

The frequency of the COI in a power system with 100~\% of \ac{GFVSC}-based generation can be defined as~\cite{Choopani2020}:
\begin{equation}\label{eq:w_COI}
\omega_{COI}=\frac{1}{H_{tot}}\sum_{k=1}^{n} H_{VSC,k} \omega_{k} \mbox{ (pu)\space, with } H_{tot}= \sum_{k=1}^{n} H_{VSC,k}.
\end{equation}

Two control strategies are proposed in this work:
\begin{itemize}
	\item Local fast voltage booster (FVB-L, for short).
	\item Fast voltage booster using a wide-area control system (WACS) (FVB-WACS, for short).
\end{itemize}

\subsection{Local fast voltage booster (FVB-L)}\label{sec:VSC_V_TS_FVB_L}
\begin{figure}[!htbp]
\begin{center}
\includegraphics[width=0.8\columnwidth]{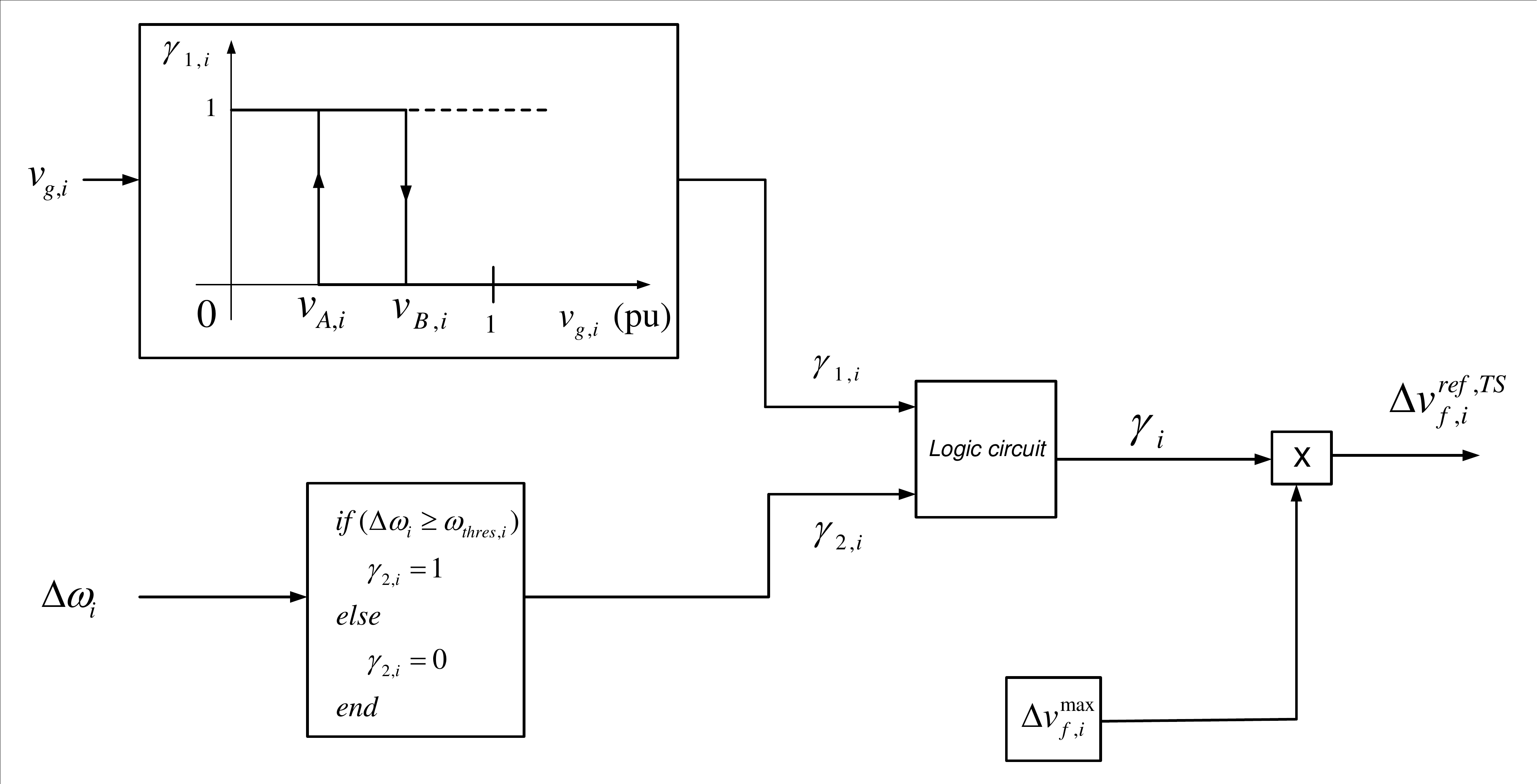}
\caption{Strategy FVB-L. }
\label{fig:VSC_GF_FVB_L}

\includegraphics[width=0.7\columnwidth]{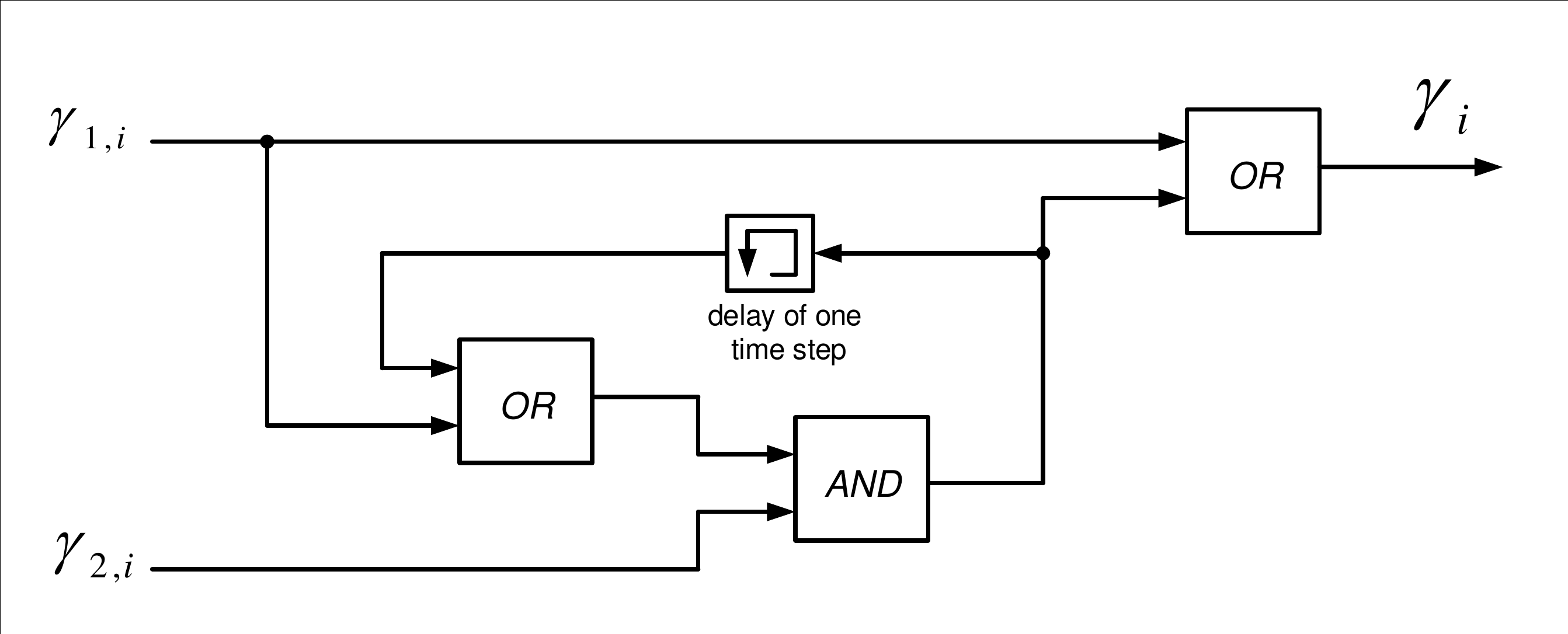}
\caption{Strategy FVB-L. Logic circuit for fault detection. }
\label{fig:VSC_GF_FVB_L_logic}
\end{center}
\end{figure}

\noindent This control strategy was motivated by previous work on excitation boosters in synchronous machines~\cite{LuisDM2016b} and on supplementary controllers for transient stability in shunt FACTS devices~\cite{Haque2004a}. It consists in a fast voltage support. Strategy FVB-L uses local measurements as input signals: the voltage at the terminal of VSC-$i$, $v_{g,i}$, and the frequency deviation of each \ac{GFVSC}-$i$, $\Delta \omega_{i}=\omega_{i}-\omega_{0,pu}$ (in pu) and its block diagram is shown in Fig.~\ref{fig:VSC_GF_FVB_L}.

The philosophy of local strategy FVB-L is as follows:
\begin{itemize}
	\item Binary variable $\gamma_{1,i}$ is set to 1 if a voltage sag is detected with an hysteresis, as shown in Fig.~\ref{fig:VSC_GF_FVB_L}. If $v_{g,i}\leq v_{A,i}$, then $\gamma_{1,i}=1$ and remains equal to 1 until $v_{g,i}>v_{B,i}$. if a fault is not detected, then $\gamma_{1,i}=0$.
	\item Binary variable $\gamma_{2,i}$ is set to 1 if the frequency deviation of \ac{GFVSC}-$i$ (with respect to the nominal frequency) is greater than or equal to a certain threshold: $\Delta \omega_{i}\ge \omega_{thres,i}$. Otherwise, $\gamma_{2,i}=0$.
	\item The supplementary controller is activated with binary variable $\gamma_{i}$, which is the result of a logic circuit with $\gamma_{1,i}$ and $\gamma_{2,i}$ as inputs, as shown in Fig.~\ref{fig:VSC_GF_FVB_L}.
	\item The supplementary voltage set point is given by: $\Delta v_{f,i}^{ref,TS}= \gamma_{i} \Delta v_{f,i}^{max}$, where $\Delta v_{f,i}^{max}>0$ is a parameter of the controller ($\gamma_{i}=0$ if the controller is deactivated and $\gamma_{i}=1$ if the controller is activated).
\end{itemize} 

The logic circuit rules for the activation of the controller can be summarised as follows:
\begin{itemize}
	\item The controller will be activated if a voltage sag is detected. Therefore, $\gamma_{1,i}$ will drive the activation of the controller.
	\item Once the controller is activated, the supplementary voltage set point is maintained if at least one of the two following conditions are satisfied: undervoltage ($\gamma_{1,i}=1$) or frequency greater than or equal to the threshold ($\gamma_{2,i}=1$).
\end{itemize}

With this controller, VSC-$i$ will increase its voltage setpoint with a positive increment $\Delta v_{f,i}^{ref,TS}=\Delta v_{f,i}^{max}$ if it detects a fault, trying to slow down the converter. This mechanism will improve transient stability of the grid-forming VSC connected to the rest of the system. The behaviour of each converter in a power system with several \acp{GFVSC} can be summarised as follows. When a fault occurs, the frequency of all \acp{GFVSC} will increase and all of them will see a certain voltage sag during the fault. The frequencies of \acp{GFVSC} close to the fault will grow faster than the frequencies of \acp{GFVSC} far from the fault. As discussed previously, the key point in transient stability is the defference between frequencies of the \acp{GFVSC} and not the absolute value of those frequencies. However, with this control strategy, converters do not have this information, because each VSC uses local measurements, only. This problem can be tackled by proper design of thresholds $v_{A,i}$, $v_{B,i}$ and $\omega_{thres,i}$ and with the logic circuit proposed of Fig.~\ref{fig:VSC_GF_FVB_L_logic}. With proper design of those thresholds, only controllers of \acp{GFVSC} close to the fault will be activated, and not the ones of VSCs far from the fault. Hence, transient stability in power systems with 100~\% of \ac{GFVSC}-based generation can be improved.

\FloatBarrier

\subsection{Fast voltage booster using a WACS (FVB-WACS)}\label{sec:VSC_V_TS_FVB_WACS}
\begin{figure}[!htbp]
\begin{center}
\includegraphics[width=1.0\columnwidth]{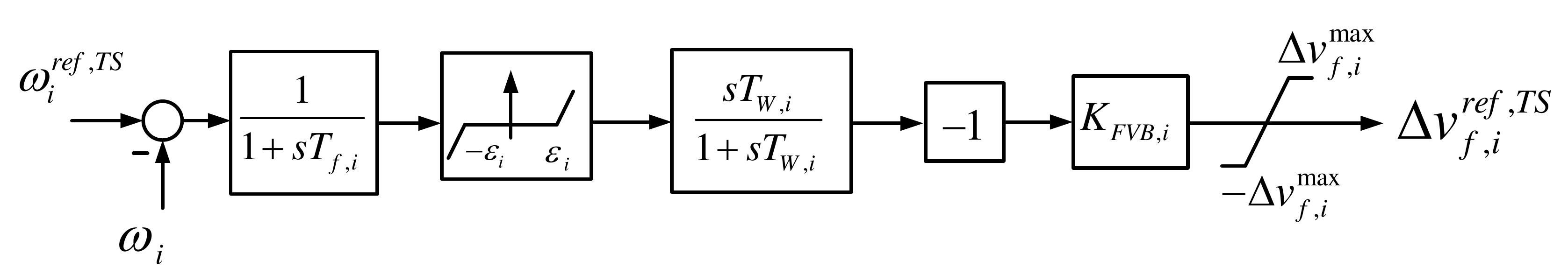}
\caption{Strategy FVB-WACS. }
\label{fig:VSC_GF_FVB_WACS}
\end{center}
\end{figure}

\noindent This control strategy was motivated by previous work on excitation boosters in synchronous machines using the speed of the COI~\cite{LuisDM2017,LuisDM2019,LuisDM2020}. The proposed controller uses a fast voltage booster in each \ac{GFVSC} using a wide-area control system (WACS). Following the scheme of Fig.~\ref{fig:VSC_GF_FVB_WACS}, where $K_{FVB,i}$ is a proportional gain, $T_{f,i}$ is a low-pass filter used for noise filtering, $T_{W,i}$ is a wash-out filter used to avoid any control actions in case of steady-state offsets and $\Delta v_{f,i}^{max}$ is a saturation parameter. The input of the controller is the error between a frequency set point ($\omega_{i}^{ref,TS}$) and the frequency of each \ac{GFVSC} ($\omega_{i}$), in pu. A deadband of $\pm \epsilon_{i}$ is used to apply the control actions only when the system is subject to large-enough perturbations.

The frequency set point of each VSC-$i$ is calculated as the frequency of the COI (Eq.~(\ref{eq:w_COI})):
\begin{equation}\label{eq:FB_WACS_w_ref_TS}
\omega_{i}^{ref,TS}=\omega_{COI}
\end{equation}
Hence, a communication system is needed.  

The philosophy of FVB-WACS controller is as follows:
\begin{itemize}
	\item If the frequency of VSC-$i$ is above the frequency of the COI, VSC-$i$ will increase its voltage setpoint, trying to slow VSC-$i$ down.
	\item If the frequency of VSC-$i$ is below the frequency of the COI, VSC-$i$ will decrease its voltage setpoint, trying to accelerate VSC-$i$.
	\item Therefore, control actions will pull together the frequencies of \acp{GFVSC} of the system.
\end{itemize}

\FloatBarrier




\newpage
\section{Results}\label{sec:Results2_VSC_V_Kundur}
\begin{figure}[!htbp]
\centering
\includegraphics[width=1.0\columnwidth]{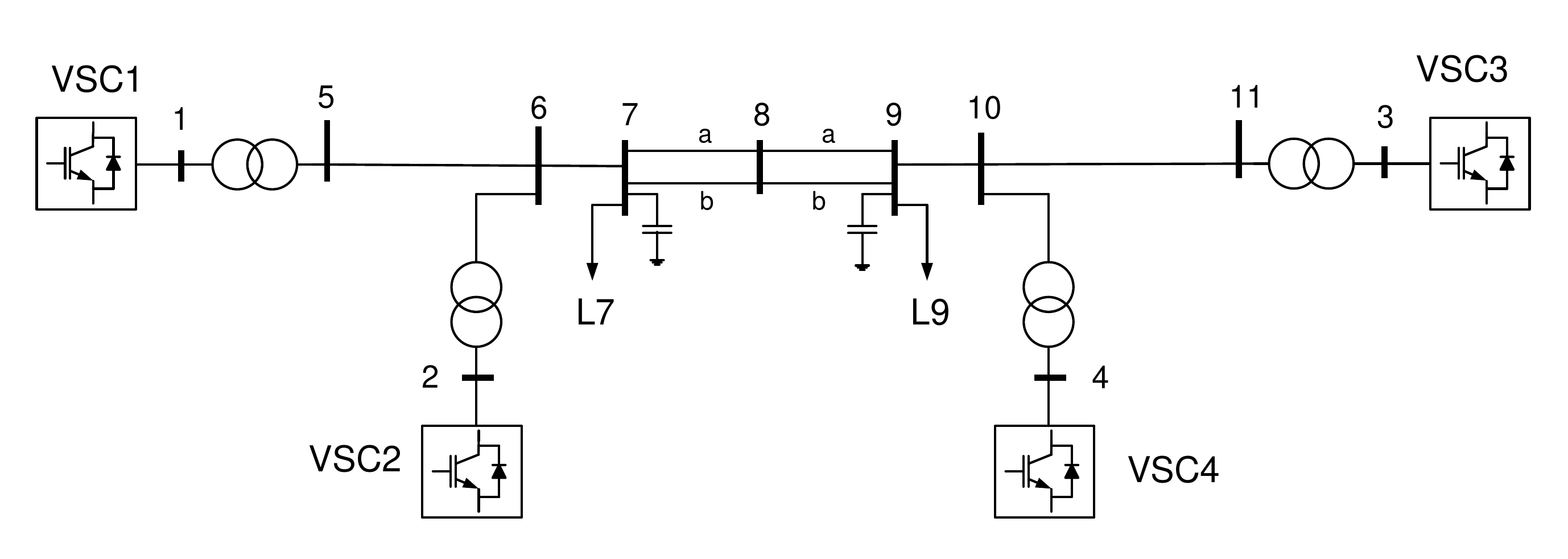}
\caption{Kundur's two-area test system with 100~\% of \ac{GFVSC}-based generation. }
\label{fig:Kundur_two_area_VSC_V}
\end{figure}

\noindent The behaviour of Kundur's two-area test system~\cite{Kundur1994} with 100\% of grid-forming VSC-based generation has been investigated (see Fig.~\ref{fig:Kundur_two_area_VSC_V}). Synchronous machines of the original system were replaced by \ac{GFVSC}-based generators with VSM control, with the same rating as the original generators (900 MVA). Data of the system are provided in the Appendix. Simulations were carried out with VSC\_Lib tool, an open-source tool based on Matlab + Simulink + SimPowerSystems developed by L2EP-LILLE~\cite{L2EP_VSC_GF_2020,MIGRATE_WP3_2018,Qoria2019}. Average electromagnetic models are used.

Table~\ref{tab:Kundur_ini_op_point} shows the results of the steady-state  initial operating point. 

\begin{table}[H]
\centering
\caption{Initial operating point.  }
\begin{tabular}{|l|cccc|}
\hline
 \textbf{VSC}  &  $v_{g,i}$ (pu)  &  $\delta_{g,i}$ (deg) &   $P_{g,i}$ (MW)  &  $Q_{g,i}$ (MVAr)            \\ 
\hline 
VSC 1& 1.0475 &  0.44  & 693.00 &  0.00     \\
VSC 2& 1.0309 &  0.45  &693.00  &  90.00     \\
VSC 3&  0.9900&  0.00 & 642.60 &   -69.93    \\
VSC 4&   0.9738  &  0.16 & 693.00 & 180.00      \\
\hline
\end{tabular}
\label{tab:Kundur_ini_op_point}
\end{table}

Three cases are compared:
\begin{itemize}
	\item Base case: no supplementary controller for transient stability is implemented in the \acp{GFVSC}.
	\item FVB-L: VSCs applying FVB-L strategy (Figs.~\ref{fig:VSC_GF_FVB_L}-\ref{fig:VSC_GF_FVB_L_logic}), with parameters: $v_{A,i}=0.75$~pu, $v_{B,i}=0.9$~pu, $\omega_{thres,i}=10^{-3}$~pu, $\Delta v_{f,i}^{max}=0.15$~pu.
	\item FVB-WACS: VSCs applying FVB-WACS strategy (Fig.~\ref{fig:VSC_GF_FVB_WACS}), with parameters: $K_{FVB,i}=50$~pu, $T_{f,i}=0.1$~s, $T_{W,i}=10$~s, $\Delta v_{f,i}^{max}=0.15$~pu and $\epsilon_{i}=10^{-3}$~pu.
\end{itemize}

\FloatBarrier
\subsection{Short-circuit simulation}\label{sec:Results2_VSC_V_Kundur_sim1}
\begin{figure}[H]
\begin{center}
\centering
\includegraphics[width=0.70\columnwidth]{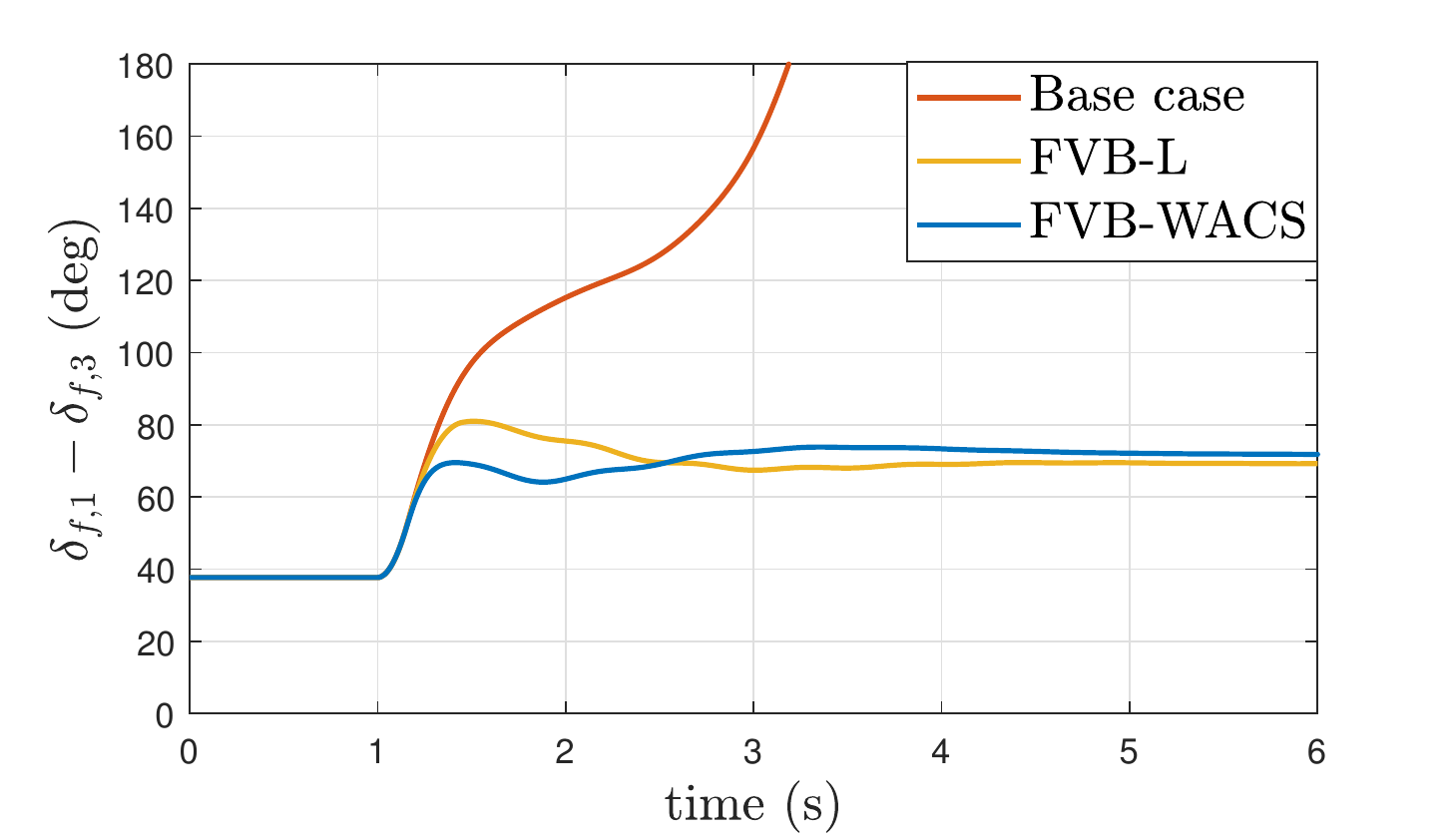}
\caption{Fault I cleared after 150~ms. Angle difference of the VSCs. }
\label{fig:Kundur_sim1_angles}
\end{center}
\end{figure}

\noindent A three-phase-to-ground short circuit is applied to line 7-8$a$ (close to bus 7), which is cleared by diconnecting the line 150~ms later (Fault I). Fig.~\ref{fig:Kundur_sim1_angles} shows the angle difference between VSC-1 and VSC-3. In the base case, VSC-based generators lose synchronism. However, synchronism is maintained with the proposed supplementary controllers FVB-L and FVB-WACS (see Fig.~\ref{fig:Kundur_sim1_angles}). 

\begin{figure}[!htbp]
\begin{center}
\centering
\includegraphics[width=0.7\columnwidth]{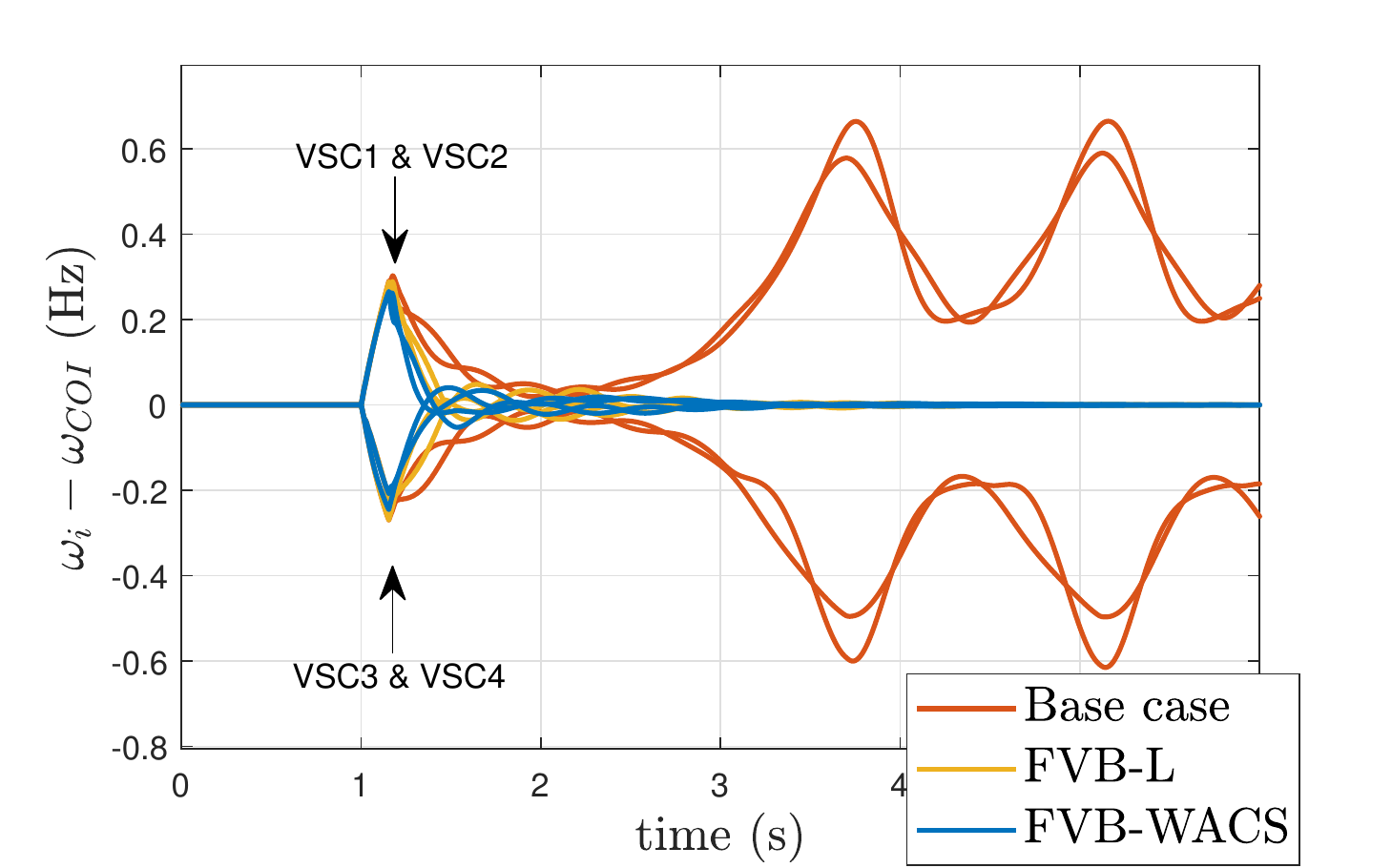}
\caption{Fault I cleared after 150~ms. Frequency deviations of the VSCs with respect to the frequency of the COI. }
\label{fig:Kundur_sim1_freq_COI}
\end{center}
\end{figure}

\begin{figure}[!htbp]
\begin{center}
\includegraphics[width=0.9\columnwidth]{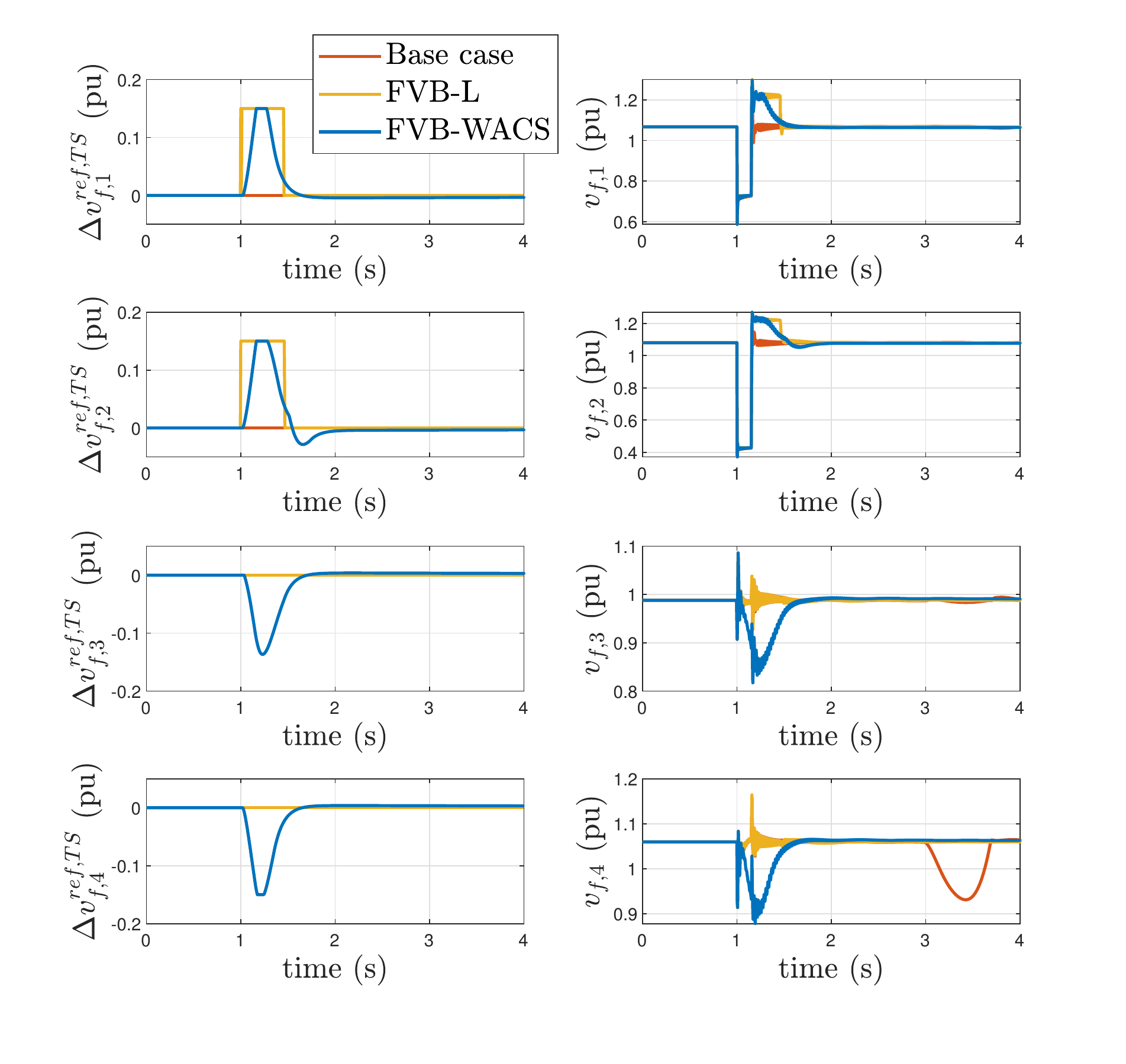}
\caption{Fault I cleared after 150~ms. (left) Supplementary voltage set points of the VSCs and (right) voltages of the VSCs. }
\label{fig:Kundur_sim1_bVf_TS}
\end{center}
\end{figure}

Fig.~\ref{fig:Kundur_sim1_freq_COI} shows the frequency deviations of the VSCs with respect to the frequency of the COI, while Fig.~\ref{fig:Kundur_sim1_bVf_TS} shows the supplementary voltage set point provided by the control strategies and the voltages of the VSCs. Notice that local strategy FVB-L is only activated in VSCs 1 and 2, which are close to the fault and not in VSCs 3 and 4, which are far from the fault. This is why strategy FVB-L is also effective in multi-converter systems with 100~\% of grid-forming based generation, and this is the consequence of the logic rules in Fig.~\ref{fig:VSC_GF_FVB_L_logic} and an appropriate design of the controller parameters. In control strategy FVB-WACS, a positive supplementary voltage set point is provided by VSCs 1 and 2 immediately after the fault clearing, because their frequencies are above the frequency of the COI (see Fig.~\ref{fig:Kundur_sim1_freq_COI}). Meanwhile, VSCs 3 and 4 provide a negative supplementary voltage during the first swing, because their frequencies are below the frequency of the COI. Therefore, VSCs 1 and 2 will slow down while VSCs 3 and 4 will accelerate, reducing the risk of loss of synchronism.

\begin{figure}[!htbp]
\begin{center}
\includegraphics[width=0.70\columnwidth]{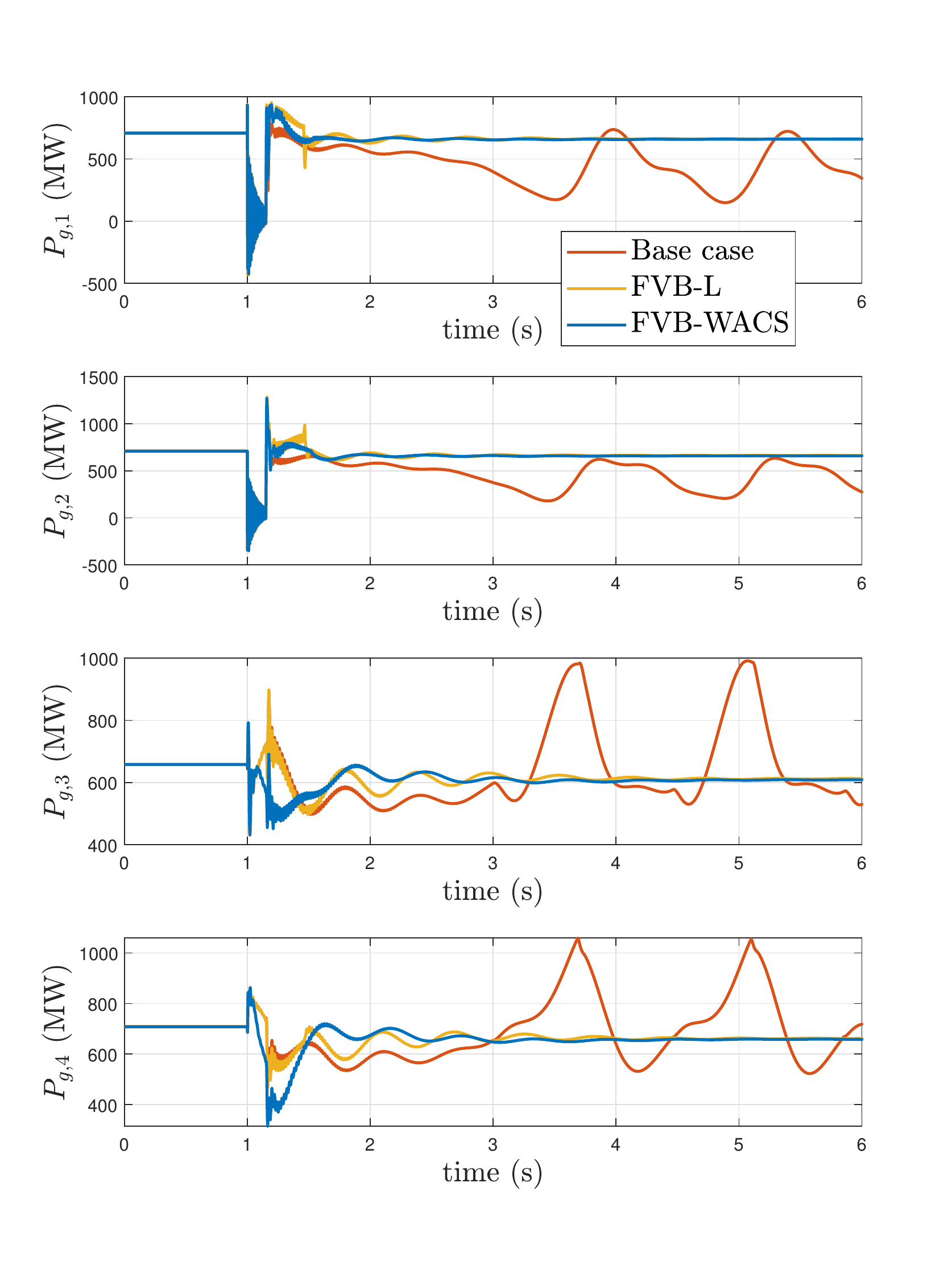}
\caption{Fault I cleared after 150~ms. Active-power injections of the VSCs. }
\label{fig:Kundur_sim1_Pg}
\end{center}
\end{figure}

Finally, Fig.~\ref{fig:Kundur_sim1_Pg} shows the active-power injections of the VSCs. During the fault, the current limiter leads to variations in the P injections, in order to maintain the current within its limit. After the fault clearing, P injections are affected by the supplementary voltage set point of the proposed control strategies (FVB-L and FVB-WACS). By supplying a positive (negative) additional voltage set point $\Delta v_{f,i}^{ref,TS}$ (Fig.~\ref{fig:Kundur_sim1_bVf_TS}), the active-power injection increases (decreases), according to~(\ref{eq:VSC_pg_electrical}). Precisely, (electrical) active-power injections, $p_{g,i}$, drive the slowing down or the acceleration of the \acp{GFVSC}, according to ~(\ref{eq_VSC_V_VSM_v1}).

\FloatBarrier
\subsection{Critical clearing times (CCTs) and impact of communication latency}\label{sec:Results2_VSC_V_Kundur_CCTs}
\noindent The critical clearing time~(CCT) of different faults (described in Table~\ref{tab:Kundur_Faults_description}) will be used to quantify transient-stability margins. Furthermore, the impact of communication latency on the performance of strategy FVB-WACS will also be analysed. Hence, the input error signal of the supplementary controller of Fig.~\ref{fig:VSC_GF_FVB_WACS} will be delayed as:
\begin{equation}\label{eq:FB_WACS_w_ref_TS_delay}
u_{i}=e^{-s \tau} (\omega_{COI}-\omega_{i})
\end{equation}
where $\tau$ is the communication delay. The work in~\cite{Chow2015} reported total communication delays in WACS within the range 50-80~ms. Total communication latency delays of 50~ms and 100~ms will be considered in this work.

CCTs of faults described in Table~\ref{tab:Kundur_Faults_description} are given in Table~\ref{tab:Kundur_CCTs}. The proposed supplementary controllers FVB-L and FVB-WACS increase the CCT of Fault I significantly. Strategy FVB-L also increases the CCT of Fault II, although the improvement is smaller than the one with strategy FVB-WACS. Furthermore, with strategy FVB-L, the CCT of Fault IV is decreased from 420~ms of the base case to 400~ms. This is due to the conditions of activation of local strategy FVB-L described in Section~\ref{sec:VSC_V_TS_FVB_L}. With a proper design of the activation thresholds, this strategy is only activated for severe-enough faults and this is why no negative impact is produced for any fault. For example, if parameter $v_{A,i}$ of Fig.~\ref{fig:VSC_GF_FVB_L} is changed from 0.75~pu to 0.5~pu, the CCT of Fault IV is not reduced (see Table~\ref{tab:Kundur_CCTs}). Besides, there could be other faults (e.g. Fault III) in which the control strategies will have no impact. In general, strategy FVB-WACS produces better results than FVB-L. Results of Table~\ref{tab:Kundur_CCTs} also prove that strategy FVB-WACS is robust against communication latency.

\begin{table}[!htbp]
\begin{center}
\caption{Fault description}
\begin{tabular}{|l|c|c|c|}
\hline
 & Short circuit  & close & clearing   \\ 
 &at line $i-j$ & to bus &   \\ \hline 
Fault I &  7-8a &  7  & Disconnect 7-8a \\  
Fault II& 5-6 &  5 & short circuit cleared  (line not disconnected)  \\
Fault III& 10-11 &  11& short circuit cleared  (line not disconnected)  \\
Fault IV &  8-9a &  8  & Disconnect 8-9a \\  
 \hline
\end{tabular}
\label{tab:Kundur_Faults_description}

\caption{Critical clearing times (CCTs).  }
\begin{tabular}{|l|c|cc|ccc|}
\hline
\textbf{CCT}   &  \textbf{base}  & \multicolumn{2}{|c|}{\textbf{FVB-L}  } & \textbf{FVB-WACS}  &       with & delay        \\ 
    (ms)              & \textbf{case}  & $v_{A,i}=0.75$~pu          & $v_{A,i}=0.5$~pu           & $\tau=0$ ms    & $50$ ms & $100$ ms    \\ \hline 
Fault I& 130  & 250 & 250 &270 & 270 & 260      \\
Fault II&  270 & 310 & 310 & 360& 340 &  320    \\
Fault III&  220 & 220 & 220  & 230 & 230 &  230     \\
Fault IV& 420 & 400 & 420 & 880 & 870 & 890      \\
\hline
\end{tabular}
\label{tab:Kundur_CCTs}
\end{center}
\end{table}


\subsection{Discussion on the use of local and global measurements}\label{sec:discussion}
\noindent This section discusses two key factors of the proposed control strategies: implementation and effectiveness. In strategy FVB-L, each \ac{GFVSC} uses local measurements, only, whereas in strategy FVB-WACS, they use global measurements and, therefore, a communication system is needed. Clearly, the implementation of local strategy FVB-L is easier and cheaper. The ideal control actions are supplying a positive (negative) supplementary voltage in those VSCs with frequency above (below) the frequency of the COI. Although strategy FVB-WACS produces better results than strategy FVB-L, precisely because it uses global measurements (the frequency of the COI), the activation thresholds of the local strategy FVB-L  (described Section~\ref{sec:VSC_V_TS_FVB_L}) can be tuned in order to ensure that the control strategy is activated only for faults that are close enough, which are, in fact, the ones that produce frequencies above the frequency of the COI.  As a general recommendation, local strategy FVB-L can be implemented in grid-forming VSCs to improve transient stability for severe faults. Nevertheless, the use of strategy FVB-WACS could be an interesting option in power systems which are vulnerable to transient stability, since the latter can produce more significant improvements.

\FloatBarrier
\newpage

\section{Conclusions}\label{sec:conclusion}
\noindent This paper has proposed two fast voltage boosters to improve transient stability of power systems with 100\% of grid-forming VSC-based generation~(\ac{GFVSC}). One is based on local measurements (FVB-L) and the other one is based on global measurements (FVB-WACS). The conclusions obtained in this paper are as follows:
\begin{itemize}
	\item Transient-stability phenomenon is present in power systems with 100\% of \ac{GFVSC}-based generation.
	\item Local strategy FVB-L improves transient stability of severe faults significantly. However, its control actions must be restricted to severe-enough faults tuning activation thresholds. This means that this control strategy will have little effect on transient stability of non-severe faults.
	\item Global strategy FVB-WACS improves transient stability, significantly. It produces significant improvements for severe and non-severe faults. The control strategy is robust when subject to communication latency.
	\item An important advantage of the two control strategies proposed is that they are based on fast voltage control, avoiding changing the set points linked to frequency/active-power injection of the grid-forming VSC~(i.e set points linked to the primary energy source of the VSC).
\end{itemize}
\vspace{-0.3cm}
\section*{Appendix: data}
\noindent Data of the grid-forming VSCs are provided in Table~\ref{tab:VSC_parameters}. Data of the original two-area Kudur's test system can be found in~\cite{Kundur1994}. Nominal voltage of the transmission grid and the nominal frequency (230~kV and 60~Hz, respectively) were changed to  220~kV and 50~Hz in this study. Loads were modelled as constant impedances for dynamic simulation. Besides, a critical case for transient stability was achieved by increasing the power transfer from Area 1 to Area 2:
\begin{itemize}
	\item Loads: bus 7: 917 MW \& 100 MVAr;  bus 9: 1817 MW \& 100 MVAr.
\end{itemize}

\begin{table}[H]
\caption{Parameters of the VSCs} 
\begin{center}
\begin{tabular}{|l|c|}
\hline
\textbf{Parameters} &   \\ 
VSC's rating are base values for pu& \\ 
\hline
Rating VSC, DC voltage, AC voltage & 900 MVA, $\pm 320$ kV, $300$ kV   \\
Max. current & 1.25 pu (equal priority for $d-q$ axes)  \\
Max. modulation index ($m_{i}^{max} = \sqrt{\frac{3}{2}} \cdot \frac{V_{dc,B}}{2 V_{ac,B}}$) &1.31 pu  \\ 
Series filter resistance ($r_{f,i}$)/reactance ($x_{f,i}$) & 0.005 pu / 0.15 pu \\
Shunt filter capacitance ($C_{f,i}$) & 0.15 pu  \\
Transformer resistance ($r_{c,i}$)/reactance ($x_{c,i}$) & 0.005 pu / 0.15 pu \\
(900 MVA 300/220 kV transformer) &  \\
Inner prop./int. control  ($K_{C,P,i}$/$K_{C,I,i}$) & 0.73 pu / 1.19 pu/s \\
Outer prop./int. control   ($K_{V,P,i}$/$K_{V,I,i}$) & 0.52 pu / 1.16 pu/s \\
Virtual transient resistance   ($r_{V,i}$/$T_{VR,i}$) & 0.09 pu / 0.0167 s \\
Emulated inertia ($H_{VSC,i}$) of VSCs 1 and 2& 4.5  s / 4.5 s  \\
 Emulated inertia ($H_{VSC,i}$) of VSCs 3 and 4 & 4.175 s / 6.175 s \\
Primary freq. controller gain. ($D_{VSC,i}$)  & 20 pu \\
\hline
\end{tabular}
\label{tab:VSC_parameters}
\end{center}
\end{table}


\section*{Acknowledgment}
Work supported  by the Spanish Government and under RETOS Project  Ref. RTI2018-098865-B-C31 (MCI/AEI/FEDER, UE) and by Madrid Regional Government under PROMINT-CM Project  Ref. S2018/EMT-4366.

\section*{Contact information of the authors}
R\'egulo E. \'Avila-Mart\'inez, Javier Renedo, Luis Rouco, Aurelio Garc\'ia-Cerrada and Lukas Sigrist are with the Instituto de Investigaci\'on Tecnol\'ogica (IIT), ETSI ICAI, 
Universidad Pontificia Comillas, Madrid, Spain (e-mail: \{regulo.avila, javier.renedo, luis.rouco, aurelio, lukas.sigrist\}@iit.comillas.edu). Taoufik Qoria and Xavier Guillaud are with the Laboratoire d'Electrotechniquede Puissance (L2EP), Centrale Lille, Arts et M\'etiers Paris Tech, University of Lille, Lille, France (e-mail: \{taoufik.qoria, xavier.guillaud\}@centralelille.fr).




\begin{thebibliography}{10}
\providecommand{\url}[1]{#1}
\csname url@samestyle\endcsname
\providecommand{\newblock}{\relax}
\providecommand{\bibinfo}[2]{#2}
\providecommand{\BIBentrySTDinterwordspacing}{\spaceskip=0pt\relax}
\providecommand{\BIBentryALTinterwordstretchfactor}{4}
\providecommand{\BIBentryALTinterwordspacing}{\spaceskip=\fontdimen2\font plus
\BIBentryALTinterwordstretchfactor\fontdimen3\font minus
  \fontdimen4\font\relax}
\providecommand{\BIBforeignlanguage}[2]{{%
\expandafter\ifx\csname l@#1\endcsname\relax
\typeout{** WARNING: IEEEtran.bst: No hyphenation pattern has been}%
\typeout{** loaded for the language `#1'. Using the pattern for}%
\typeout{** the default language instead.}%
\else
\language=\csname l@#1\endcsname
\fi
#2}}
\providecommand{\BIBdecl}{\relax}
\BIBdecl

\bibitem{Rocabert2012}
J.~Rocabert, A.~Luna, F.~Blaabjerg, and P.~Rodr\'{i}guez, ``{Control of Power
  Converters in AC Microgrids},'' \emph{IEEE Transactions on Power
  Electronics}, vol.~27, no.~11, pp. 4734--4749, 2012.

\bibitem{Guerrero2013}
J.~M. Guerrero, M.~Chandorkar, T.-L. Lee, and P.-C. Loh, ``{Advanced Control
  Architectures for Intelligent Microgrids - Part I: Decentralized and
  Hierarchical Control},'' \emph{IEEE Transactions on Industrial Electronics},
  vol.~60, no.~4, pp. 1254--1262, 2013.

\bibitem{Olivares2014}
D.~E. Olivares, A.~Mehrizi-Sani, A.~H. Etemadi, C.~A. Canizares, R.~Iravani,
  M.~Kazerani, A.~H. Hajimiragha, O.~Gomis-Bellmunt, M.~Seedifard,
  R.~Palma-Behnke, A.~Jim\'{e}nez-Est\'{e}vez, and N.~Hatziargyriou, ``{Trends
  in Microgrid Control},'' \emph{IEEE Transactions on Smart Grid}, vol.~5,
  no.~4, pp. 1905--1919, 2014.

\bibitem{Milano_lowH2018}
F.~Milano, F.~D$\ddot{o}$rfler, G.~Hug, D.~J. Hill, and Verbi$\check{c}$,
  ``{Foundations and Challenges of Low-Inertia Systems},'' in \emph{Proc. Power
  Systems Computation Conference (PSCC), Dublin, Ireland}, 2018, pp. 1--25.

\bibitem{Paolone2020}
M.~Paolone, T.~Gaunt, X.~Guillaud, M.~Liserre, S.~Meliopoulos, A.~Monti,
  T.~{Van Cutsen}, V.~Vittal, and C.~Vournas, ``{Fundamentals of power systems
  modelling in the presence of converter-interfaced generation},''
  \emph{Electric Power Systems Research}, vol. 189, no. 106811, pp. 1--33,
  2020.

\bibitem{Zhong2011}
Q.-C. Zhong and G.~Weiss, ``{Synchronverters: Inverters That Mimic Synchronous
  Generators},'' \emph{IEEE Transactions on Industrial Electronics}, vol.~58,
  no.~4, pp. 1259--1267, 2011.

\bibitem{DArco2015}
S.~{D'Arco}, J.~A. Suul, and O.~B. Fosso, ``{A Virtual Synchronous Machine
  implementation for distributed control of power converters in SmartGrids},''
  \emph{Electric Power Systems Research}, vol. 122, pp. 180--197, 2015.

\bibitem{jroldan2019}
J.~Rold\'an-P\'erez, A.~Rodr\'iguez-Cabero, and M.~Prodanovic, ``{Design and
  analysis of virtual synchronous machines in inductive and resistive weak
  grids},'' \emph{IEEE Transactions on Energy Conversion}, vol.~34, no.~2, pp.
  1818--1828, 2019.

\bibitem{DArco2014}
S.~{D'Arco} and J.~A. Suul, ``{Equivalence of Virtual Synchronous Machines and
  Frequency-Droops for Converter-Based Microgrids},'' \emph{IEEE Transactions
  on Smart Grid}, vol.~5, no.~1, pp. 394--395, 2014.

\bibitem{Amenedo2021}
J.~L. {Rodr\'iguez Amenedo}, S.~{Arnaltes G\'omez}, and M.~Alonso-Mart\'inez,
  J~{Gonz\'alez de Armas}, ``{Grid-Forming Converters Control Based on the
  Reactive Power Synchronization Method for Renewable Power Plants},''
  \emph{IEEE Access}, no.~9, pp. 67\,989--68\,007, 2021.

\bibitem{Nikos2020}
N.~Hatziargyriou, J.~V. Milanovi\'{c}, C.~Rahmann, V.~Ajjarapu, C.~Canizares,
  I.~Erlich, D.~Hill, I.~Hiskens, I.~Kamwa, B.~Pal, P.~Pourbeik, J.~J.
  Sanchez-Gasca, A.~Stankovi\'{c}, T.~{Van Cutsem}, V.~Vittal, and C.~Vournas,
  ``{Definition and Classification of Power System Stability - Revisited and
  Extended},'' \emph{IEEE Transactions on Power Systems}, vol. doi:
  10.1109/TPWRS.2020.3041774, pp. 1--12, 2020.

\bibitem{Andrade2011}
F.~Andrade, J.~Cusid\'{o}, and L.~Romeral, ``{Transient stability analysis of
  inverter-interfaced distributed generators in a microgrid system},'' in
  \emph{Proc. European Conference on Power Electronics and Applications,
  Birmingham, UK}, 2011, pp. 1--10.

\bibitem{Andrade2011b}
F.~Andrade, J.~Cusid\'{o}, L.~Romeral, and J.~J. C\'{a}rdenas, ``{Study of
  transient stability for parallel connected inverters in Microgrid system
  works in stand-alone},'' in \emph{Proc. World Renewable Energy Congress,
  BLinköping, Sweden}, 2011, pp. 1--8.

\bibitem{Xin2016}
H.~Xin, L.~Huang, L.~Zhang, Z.~Wang, and J.~Hu, ``{Synchronous Instability
  Mechanism of P-f Droop-Controlled Voltage Source Converter Caused by Current
  Saturation},'' \emph{IEEE Transactions on Power Systems}, vol.~31, no.~6, pp.
  5206--5207, 2016.

\bibitem{Eskandari2020}
M.~Eskandari and A.~V. Savkin, ``{On the Impact of Fault Ride-Through on
  Transient Stability of Autonomous Microgrids: Nonlinear Analysis and
  Solution},'' \emph{IEEE Transactions on Smart Grid}, vol. doi:
  10.1109/TSG.2020.3030015, pp. 1--12, 2020.

\bibitem{Shuai2018}
Z.~Shuai, C.~Shen, X.~Liu, Z.~Li, and Z.~J. Shen, ``{Transient Angle Stability
  of Virtual Synchronous Generators Using Lyapunov’s Direct Method},''
  \emph{IEEE Transactions on Smart Grid}, vol.~10, no.~4, pp. 4648--4661, 2018.

\bibitem{Cheng2020}
H.~Cheng, Z.~Shuai, C.~Shen, X.~Liu, Z.~Li, and Z.~J. Shen, ``{Transient Angle
  Stability of Paralleled Synchronous and Virtual Synchronous Generators in
  Islanded Microgrids},'' \emph{IEEE Transactions on Power Electronics},
  vol.~35, no.~8, pp. 1019--1033, 2020.

\bibitem{Pan2020}
D.~Pan, X.~Wang, F.~Liu, and R.~Shi, ``{Transient Stability of Voltage-Source
  Converters With Grid-Forming Control: A Design-Oriented Study},'' \emph{IEEE
  Journal of Emerging and Selected Topics in Power Electronics}, vol.~8, no.~2,
  pp. 1019--1033, 2020.

\bibitem{Qoria_VSC_current_limit2020}
T.~Qoria, F.~Gruson, F.~Colas, X.~Kestelyn, and X.~Guillaud, ``{Current
  limiting algorithms and transient stability analysis of grid-forming VSCs},''
  \emph{Electric Power Systems Research}, vol. 189, no. 106726, pp. 1--8, 2020.

\bibitem{Qoria_VSC_CCT2020}
T.~Qoria, F.~Gruson, F.~Colas, G.~Denis, T.~Prevost, and X.~Guillaud,
  ``{Critical Clearing Time Determination and Enhancement of Grid-Forming
  Converters Embedding Virtual Impedance as Current Limitation Algorithm},''
  \emph{IEEE Journal of Emerging and Selected Topics in Power Electronics},
  vol.~8, no.~2, pp. 1050--1061, 2020.

\bibitem{Qoria2020}
T.~Qoria, E.~Rokrok, A.~Bruyere, B.~Francois, and X.~Guillaud, ``{A PLL-Free
  Grid-Forming Control With Decoupled Functionalities for High-Power
  Transmission System Applications},'' \emph{IEEE Access}, no. 106765, pp.
  197\,363--197\,378, 2020.

\bibitem{Shen2020}
C.~Shen, Z.~Shuai, Y.~Shen, Y.~Peng, X.~Liu, Z.~Li, and J.~Shen, ``{Transient
  Stability and Current Injection Design of Paralleled Current-Controlled VSCs
  and Virtual Synchronous Generators},'' \emph{IEEE Transactions on Smart
  Grid}, vol. doi: 10.1109/TSG.2020.3032610, pp. 1--15, 2020.

\bibitem{Choopani2020}
M.~Choopani, S.~H. Hosseinian, and B.~Vahidi, ``{New Transient Stability and
  LVRT Improvement of Multi-VSG Grids Using the Frequency of the Center of
  Inertia},'' \emph{IEEE Transactions on Power Systems}, vol.~35, no.~1, pp.
  527--538, 2020.

\bibitem{Xiong2021}
X.~Xiong, C.~Wu, P.~Cheng, and F.~Blaabjerg, ``{An Optimal Damping Design of
  Virtual Synchronous Generators for Transient Stability Enhancement},''
  \emph{IEEE Transactions on Power Electronics}, vol. doi:
  10.1109/TPEL.2021.3074027, pp. 1--5, 2021.

\bibitem{Lee1986}
D.~C. Lee and P.~Kundur, ``Advanced excitation controls for power system
  stability enhancement,'' in \emph{CIGRE}, no. 38-01, 1986.

\bibitem{Kundur1994}
P.~Kundur, \emph{Power System Stability and Control}.\hskip 1em plus 0.5em
  minus 0.4em\relax McGraw Hill Education, 1994.

\bibitem{LuisDM2016b}
L.~D\'{i}ez-Maroto, L.~Rouco, and F.~Fern\'{a}ndez-Bernal, ``{ Modeling, sizing
  and control of an excitation booster for enhancement of synchronous
  generators fault ride through capability: experimental validation},''
  \emph{IEEE Transactions on Energy Conversion}, vol.~31, no.~4, pp.
  1304--1314, 2019.

\bibitem{LuisDM2017}
L.~Diez-Maroto, L.~Vanfretti, M.~S. Almas, G.~M. J\'{o}nsd\'{o}ttir, and
  L.~Rouco, ``{A WACS exploiting generator Excitation Boosters for power system
  transient stability enhancement},'' \emph{Electric Power Systems Research},
  vol. 148, pp. 245--253, 2017.

\bibitem{LuisDM2019}
L.~D\'{i}ez-Maroto, J.~Renedo, L.~Rouco, and F.~Fern\'{a}ndez-Bernal,
  ``{Lyapunov Stability Based Wide Area Control Systems for Excitation Boosters
  in Synchronous Generators},'' \emph{IEEE Transactions on Power Systems},
  vol.~34, no.~1, pp. 194--204, 2019.

\bibitem{LuisDM2020}
------, ``{Wide area controllers for excitation boosters for transient
  stability improvement},'' \emph{Electric Power Systems Research}, vol. 189,
  no. 106622, pp. 1--6, 2020.

\bibitem{Haque2004a}
M.~H. Haque, ``{Improvement of First Swing Stability Limit by Utilizing Full
  Benefit of Shunt FACTS Devices},'' \emph{IEEE Transactions on Power Systems},
  vol.~19, no.~4, pp. 1894--1902, 2004.

\bibitem{Fuchs2014}
A.~Fuchs, M.~Imhof, T.~Demiray, and M.~Morari, ``{Stabilization of Large Power
  Systems Using VSC-HVDC and Model Predictive Control},'' \emph{IEEE
  Transactions on Power Delivery}, vol.~29, no.~1, pp. 480--488, 2014.

\bibitem{iitcontrolQ2017}
J.~Renedo, A.~Garc\'ia-Cerrada, and L.~Rouco, ``{Reactive-Power Coordination in
  VSC-HVDC Multi-Terminal Systems for Transient Stability Improvement},''
  \emph{IEEE Transactions on Power Systems}, vol.~32, no.~5, pp. 3758--3767,
  2017.

\bibitem{Qoria2018}
T.~Qoria, F.~Gruson, F.~Colas, X.~Guillaud, M.-S. Debry, and T.~Prevost,
  ``{Tuning of cascaded controllers for robust grid-forming Voltage Source
  Converter},'' in \emph{Proc. Power Systems Computation Conference (PSCC),
  Dublin, Ireland}, 2018, pp. 1--8.

\bibitem{Rokrok2020}
E.~Rokrok, T.~Qoria, A.~Bruyere, B.~Francois, and X.~Guillaud,
  ``{Classification and dynamic assessment of droop-based grid-forming control
  schemes: Application in HVDC systems},'' \emph{Electric Power Systems
  Research}, vol. 189, no. 106765, pp. 1--7, 2020.

\bibitem{Pereira2020}
G.~S. Pereira, V.~Costan, A.~Bruy\'{e}re, and X.~Guillaud, ``{Simplified
  approach for frequency dynamics assessment of 100\% power electronics-based
  systems},'' \emph{Electric Power Systems Research}, vol. 188, no. 106551, pp.
  1--8, 2020.

\bibitem{L2EP_VSC_GF_2020}
L2EP-LILLE, ``{VSC\_Lib: Grid Forming Models for Matlab/SimPowerSystem},'' vol.
  https://github.com/l2ep-epmlab/ (accessed 08-07-2020), 2020.

\bibitem{MIGRATE_WP3_2018}
T.~Qoria, Q.~Cossart, C.~Li, X.~Guillaud, F.~Colas, F.~Gruson, and X.~Kestelyn,
  ``{WP3-Control and Operation of a Grid with 100\% Converter-Based Devices.
  Deliverable 3.2: Local control and simulation tools for large transmission
  systems},'' MIGRATE Project, Tech. Rep., 2018.

\bibitem{Qoria2019}
T.~Qoria, F.~Gruson, F.~Colas, G.~Denis, T.~Prevost, and X.~Guillaud,
  ``{Inertia effect and load sharing capability of grid forming converters
  connected to a transmission grid},'' in \emph{Proc. 15th IET International
  Conference on AC and DC Power Transmission (ACDC), Coventry, UK}, 2019, pp.
  1--6.

\bibitem{Chow2015}
F.~Zhang, Y.~Sun, L.~Cheng, X.~Li, J.~H. Chow, and W.~Zhao, ``{Measurement and
  Modeling of Delays in Wide-Area Closed-Loop Control Systems},'' \emph{IEEE
  Transactions on Power Systems}, vol.~30, no.~1, pp. 2426--2433, 2015.

\end{thebibliography}
\end{document}